\documentclass{osa-article}

\usepackage{amsmath}

\journal{osajournal}

\articletype{Research Article}

\usepackage{lineno}

\begin{document}

\title{Laboratory Demonstration of Image-Plane Self-Calibration in Interferometry}

\author{Christopher L. Carilli,\authormark{1,*} Bojan Nikolic,\authormark{2} Laura Torino,\authormark{3} Ubaldo Iriso,\authormark{3} and Nithyanandan Thyagarajan\authormark{4} }

\address{\authormark{1}National Radio Astronomy Observatory, P. O. Box 0, Socorro, NM 87801, US\\
\authormark{2}Astrophysics Group, Cavendish Laboratory, University of Cambridge, Cambridge CB3 0HE, UK\\
\authormark{3}ALBA - CELLS Synchrotron Radiation Facility\\Carrer de la Llum 2-26, 08290 Cerdanyola del Vallès (Barcelona), Spain\\
\authormark{4}Commonwealth Scientific and Industrial Research Organisation (CSIRO), Space \& Astronomy, P. O. Box 1130, Bentley, WA 6102, Australia
}

\email{\authormark{*}ccarilli@nrao.edu} 



\begin{abstract}
We demonstrate the Shape-Orientation-Size conservation principle for a 3-element interferometer using aperture plane masking at the ALBA visible synchrotron radiation light source. We then use these data to demonstrate Image Plane Self-Calibration. 
\end{abstract}


\section{Introduction} \label{sec:intro}

In \cite{CNT2023} we presented the concept of image plane self-calibration (IPSC) in interferometry, and demonstrated the concept with numerical simulations of interferometric data. In this follow-up report, we present a laboratory interferometric demonstration of the IPSC process. The data were taken using the Xanadu visible light diagnostics beamline at the ALBA synchrotron light source \cite{Torino2016}. Xanadu at ALBA is an ideal facility to explore various aspects of interferometry, including the effects of redundant sampling of interferometric spacings, shape-orientation-size conservation for three apertures \cite{Thyagarajan+Carilli2022}, and image plane self-calibration \cite{Carilli2024, CNT2023}. 

The basics of the interferometric concepts, including the definition of, and equations for, a visibility, closure phase, and IPSC can be found in \cite{CNT2023, Thyagarajan+Carilli2022}. In brief, interferometric imaging entails measuring the cross-correlation, or coherence, between the electromagnetic voltages measured at spatially distinct locations ('apertures', or 'interferometric elements'). The Fourier transform of these coherences, also known as visibilities, corresponds to the intensity distribution of light from the distant source \cite{vanCittert34,Zernike38,TMS2017}. The measurements in the Fourier domain are referred to as 'uv-data points' or 'visibilities', where the u,v coordinates are the vector baselines defined by the distance between holes, or elements, in the interferometric mask, measured in wavelengths. It is often assumed, and experimentally designed, that the the u,v-data are measured on a planar surface, although non-planar visibility measurements still obey a 3-dimensional Fourier transform relationship with source structure. 

Closure phase is the argument of the triple product of complex visibilities on a closed triad of inteferometric elements, or apertures. Closure phase has the important property of being robust to phase corruptions that can be factored into individual element-based terms \cite{Jennison1958}. \cite{Thyagarajan+Carilli2022}  present a geometric understanding of how closure phase manifests itself in the image plane. In essence, the shape, orientation, and size (SOS) of the image-plane triangle enclosed by the fringes of a three element interferogram, are invariant to element-based phase errors, with the only degree of freedom being an unknown translation of the grid pattern of triangles due to these phase errors. 

A straight-forward means of visualizing how SOS conservation works is given in Figure 4 in \cite{Thyagarajan+Carilli2022}: for any three element interferometer, the only possible image corruption due to an element-based phase screen is a tilt of the aperture plane, leading to a shift in the image plane. No higher order decoherence or image blurring is possible, since three points always define a plane parallel to which the wavefronts are coherent. SOS conservation is not true for an interferogram made with four or more elements, since multiple phase-planes can occur for different closed triads of interferometric elements, and higher order decoherence occurs (ie. image blurring). 

SOS conservation then raises the possibility of an image-plane self-calibration process (IPSC; \cite{CNT2023}). The process employs a model source image, plus the set of triad images made from all the three element interferometers that can be defined in the array. These three element images are cross correlated with the model source image, to derive the unknown, and idiosyncratic, shifts in the two-dimensional source coordinates due to element-based phase errors, such that the peak in the cross correlation corresponds to the required shift for each triad image to restore the true source position. The shifted triad images are then summed, leading to a more coherent image of the source brightness distribution, which can be used as a model for further iteration of the process. 

\cite{CNT2023} use numerical simulations of interferometric data to demonstrate the SOS principle and image plane self-calibration for simple sources. In this follow-up report, we employ laboratory interferometric data to both demonstrate the SOS principle, and for a laboratory demonstration of IPSC. The experimental set-up is simple, with only a single triad employed in the IPSC process, and hence detailed imaging of source structure is not possible due to the very sparse uv-coverage. However, the data are adequate to show a clear improvement in visibility coherence with IPSC iteration, consistent with the known structure of the emitting source. 


\section{Visible light Synchrotron Radiation Interferometry at ALBA}

The details of the visibile light interferometric set-up at the ALBA synchrotron light source are given in \cite{Torino2016, Nikolic2024, Carilli2024}. These include aperture mask location, reimaging optics to achieve far-field equivalence for the mask, narrow band filters centered at 538~nm with a bandwidth of 10~nm, and CCD camera imaging. The distance from the mask to the target source, which is used to relate angular size measurements to physical size of the electron beam, was 15.05~m. 

\begin{figure}[!htb]
\centering 
\centerline{\includegraphics[scale=0.3]{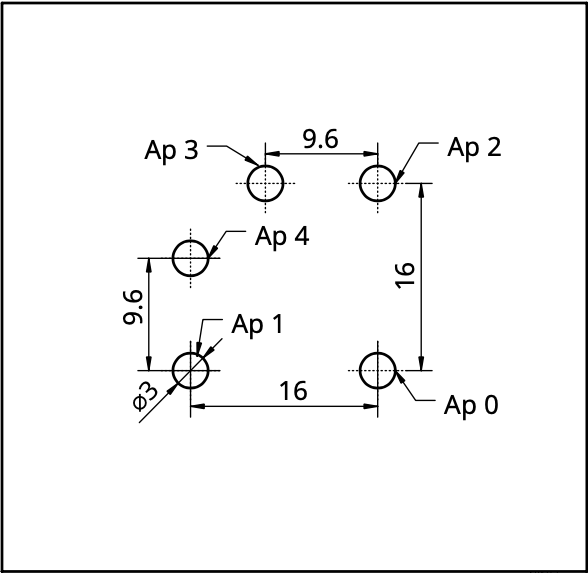}}
\caption{Scale drawing of the 5-hole mask with aperture labels as used for labelling uv-sample baselines. Dimensions are in mm (see \cite{Nikolic2024}.}
\label{fig:mask6H}
\end{figure}

Herein, we present results from a three hole mask, and from a non-redundant five hole mask, each with hole diameters of 3~mm. For reference, the geometry of our 5-hole mask is shown schematically in Figure~\ref{fig:mask6H}. Subsets of holes are covered to achieve different masks. Holes Ap 0, 1, and 2 are used for the three hole interferometry, and holes 3 and 4 are added for the 5-hole demonstration. For each mask, a series of 30 CCD frames are recorded, each of 1~ms duration, and separated by 1~s.  

The emitting source is the relativistic electron beam of the ALBA accelerator, at the location of one of the primary bending magnets. The size is small, corresponding to a Gaussian with major axis dispersion $\sim 60\mu m$ \cite{Nikolic2024}, which at a distance of 15.05\,m implies an angular size of $0.84"$. For comparison, the angular interferometric fringe spacing of our longest baseline in the mask of 22.6\,mm at 540~nm wavelength is $5"$. Hence, for all of our measurements, the visibility coherences are high, $\ge 65\%$. However, the signal to noise is extremely high, with millions of photons in each measurement, thereby allowing size measurements on partially resolving baselines. Further, for the IPSC demonstration herein, the small-scale details of the source structure are not critical. 

Visibility phases and amplitudes are derived by a Fourier transform of the interferograms, as described in detail in \cite{Carilli2024, Nikolic2024}. In the context of a study of image plane self-calibration, we first must demonstrate that any phase corruption of the visibilities can be factored into element-based phases, otherwise closure and IPSC no longer apply. 

\begin{figure}[!htb]
\centering 
\centerline{\includegraphics[scale=0.6]{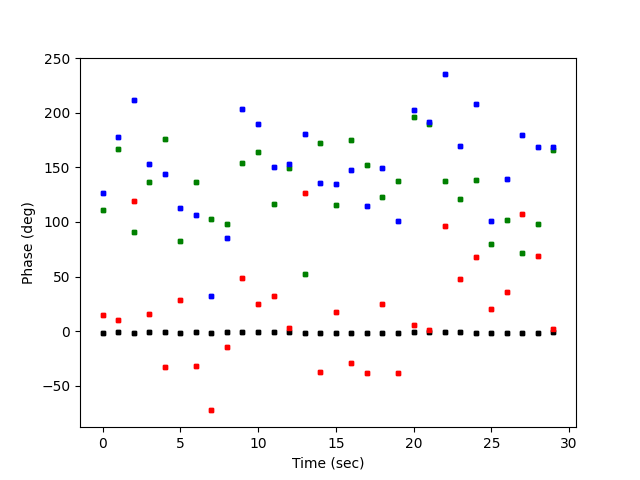}}
\caption{Visibility phases (color) and the closure phase (black) for the three hole data time series without centering, for baselines: 0-1 (green), 0-2 (red), 1-2 (blue).
}
\label{fig:CP3H}
\end{figure}

The results for visibility phases and the closure phase for the time series of the 30 visibilities for the 3-hole mask, are plotted in Figure~\ref{fig:CP3H}. This figure shows large phase variations over the time series for the visibility phases, with  rms variations of $\ge 30^o$. However, the closure phase, corresponding to the complex sum of the three phases in the closed triad, shows very small scatter (rms = $0.3^o$), and a mean value close to zero. The very small scatter in the closure phase, as compared to the visibility phases out of which it was calculated, implies that indeed, for the ALBA SRI experiment, the phase fluctuations can be factored into element-based terms. The origin of these phase fluctuations at the ALBA SRI are under investigation, but likely result from vibration of optical components, with a possible contribution from turbulence in the laboratory atmosphere. 

The near zero closure phase is principally due to the very small source size relative to the resolution of the instrument, although the source is known intrinsically to be a Gaussian shape \cite{Sands, Elleaume:km0006}, which would also have a zero closure phase regardless of resolution, since it is point-symmetric in the image plane. 

\section{Laboratory IPSC}
\label{sec:IPSC}

\subsection{Movie of Shape-Orientation-Size Conservation}

The image plane self-calibration process relies on the SOS principle for a closed triad of baselines, and the fact that the only degree of freedom due to element-based phase errors for a three element interferometer is a rigid shift of the interferogram. Again, IPSC does not work for a mask with more than three holes, since the phase screen may have structure that defines multiple phase-planes for different triads across the interferometric array, or mask \cite{CNT2023}.  It also does not work for two hole interferometry since self-calibration of an individual fringe of different orientation or baseline length is under-constrained, and the self-calibration process simply turns the source into the model. 

\begin{figure}[!htb]
\centering 
\centerline{\includegraphics[scale=0.2]{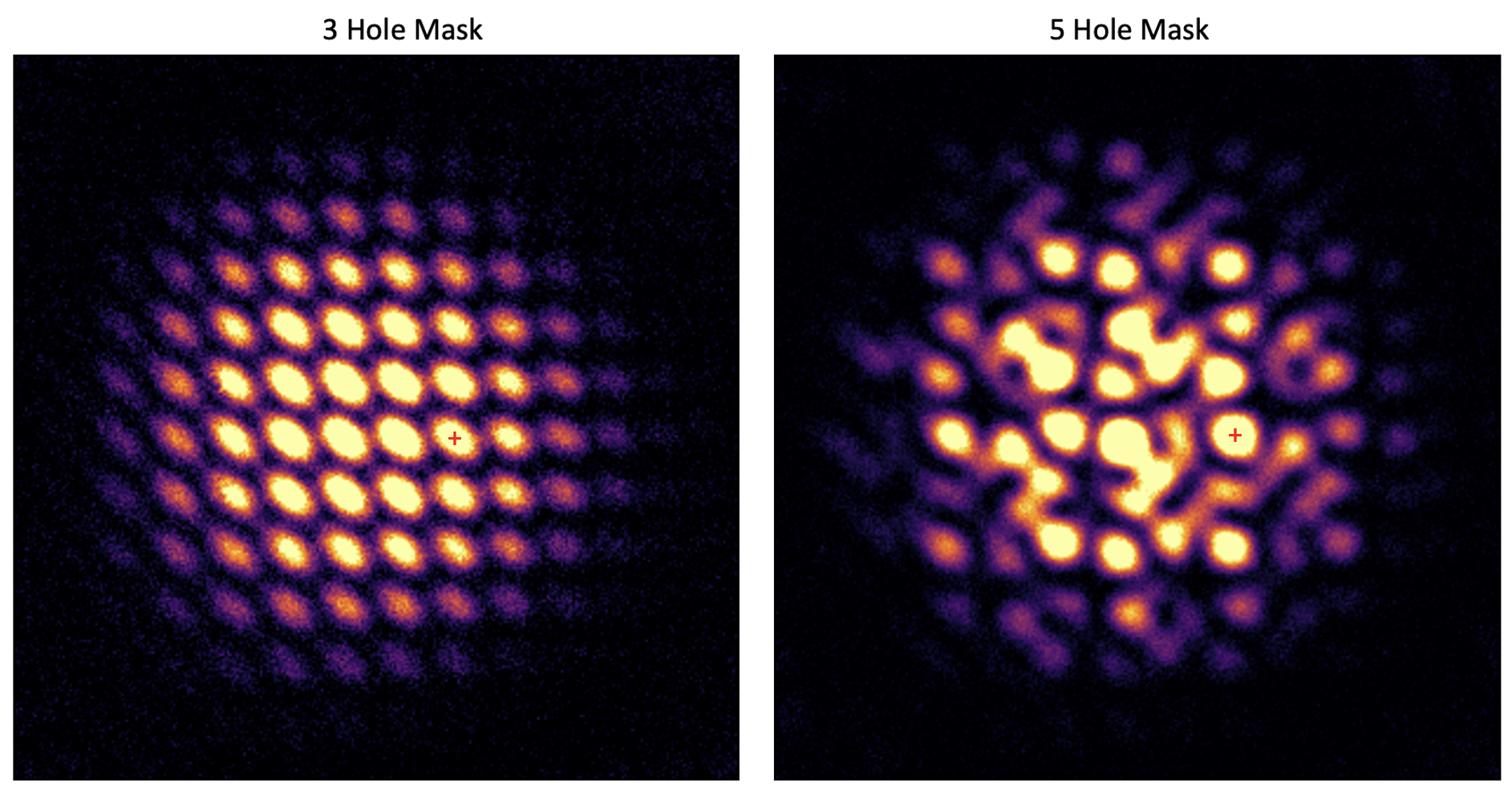}}
\caption{\url{https://www.aoc.nrao.edu/~ccarilli/TALKS/SOSMovie.pptx}. Left is the movie of the time series of interferograms for the 3-hole mask, while right is the same for the 5-hole mask. This will appear as a video in the on-line version of the Journal. The red crosses indicate the local maxima employed in Figure~\ref{fig:peaks}. The color scale is modestly saturated for the higher peaks to show lower surface brightness structures.
}
\label{fig:movie}
\end{figure}

A laboratory visualization of the SOS principle for the ALBA data can be seen in the movie of the time series of the 3-hole and 5-hole interferograms in Figure~\ref{fig:movie}. The 3-hole data shows the characteristic grid of sources resulting from such a three fringe image, multiplied by the large-scale power pattern of the individual holes, which, for circular apertures, corresponds to an Airy-disk of diameter proportional to the inverse of the hole diameter. This grid pattern undergoes rigid shifts from frame to frame due to element based phase terms, while maintaining the shape, orientation, and size of the image pattern. 

The 5-hole interferogram is more complex than the 3-hole interferogram due to the increased number of uv-sample baselines (10). The 5-hole interferogram also shows jitter in position of the overall pattern, relating to the lowest order term in the phase perturbation power spectrum, corresponding to a tip-tilt, or image shift. However, the detailed pattern of the images also changes from frame to frame, with maxima and minima increasing, decreasing, and shifting independently. This behaviour is indicative of a phase screen that has variable structure across the mask, beyond a simple gradient. SOS is not conserved and fringes from different uv-baselines cohere or decohere at a given location, depending on their relative phases at a given time.

\begin{figure}[!htb]
\centering 
\centerline{\includegraphics[scale=0.65]{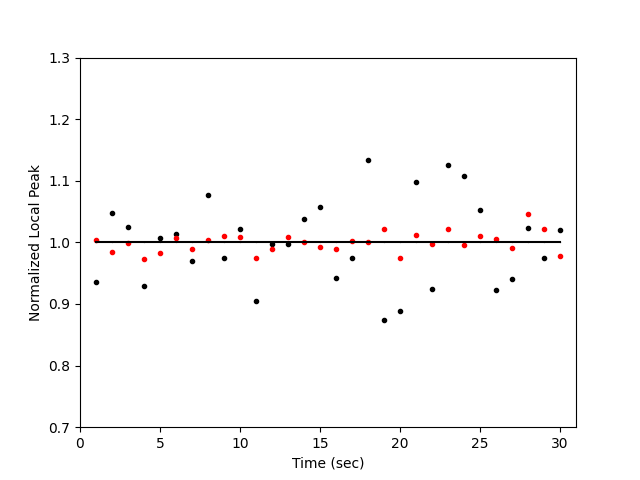}}
\caption{The time series of values of the local maximum for a well defined and isolated peak in one of the interferometric images. The black points are for the 5-hole mask, and the red points for the 3-hole mask. Values have been normalized to the mean in each case.}
\label{fig:peaks}
\end{figure}

This lack of SOS conservation for a 5-hole mask is quantified in Figure~\ref{fig:peaks}. This figure shows the time series of the local maximum for a well defined and isolated peak in the 5-hole and 3-hole interferograms. A region was chosen to have a peak of about 2/3 of the maximum on the image in each case (see Figure~\ref{fig:movie}). The photon noise per pixel necessitates a parametric fitting (in this case, using a Gaussian shape), to interpolate for the local maximum. In each time frame, the fitted region was adjusted to be centered on the feature (although exact centering made essentially no difference). For comparative purposes, all values in a given time series are normalized by the mean of the time series.

For the 3-hole images, the peak is very stable, with an rms variation of 1.6\%, likely due to pixelization noise. For the 5-hole mask, the peak shows much larger time variation, with an rms of 6.9\%. This larger variation is a quantitative reflection of the visually apparent time variation of the image pattern seen in the 5-hole movie in Figure~\ref{fig:movie}, in particular, the variation in coherence for a given feature over time. 

\subsection{Three Hole Interferograms}

We cannot use our 5-hole interferograms to demonstrate IPSC since all 10 visibility fringes are obtained in a single CCD frame, and hence individual triad images cannot be separated in the image-plane. While the 3-hole interferogram provides very limited imaging capability, the 3-hole measurements can still be used to demonstrate improved visibility coherence with IPSC in the following manner.

\begin{figure}[!htb]
\centering 
\centerline{\includegraphics[scale=0.3]{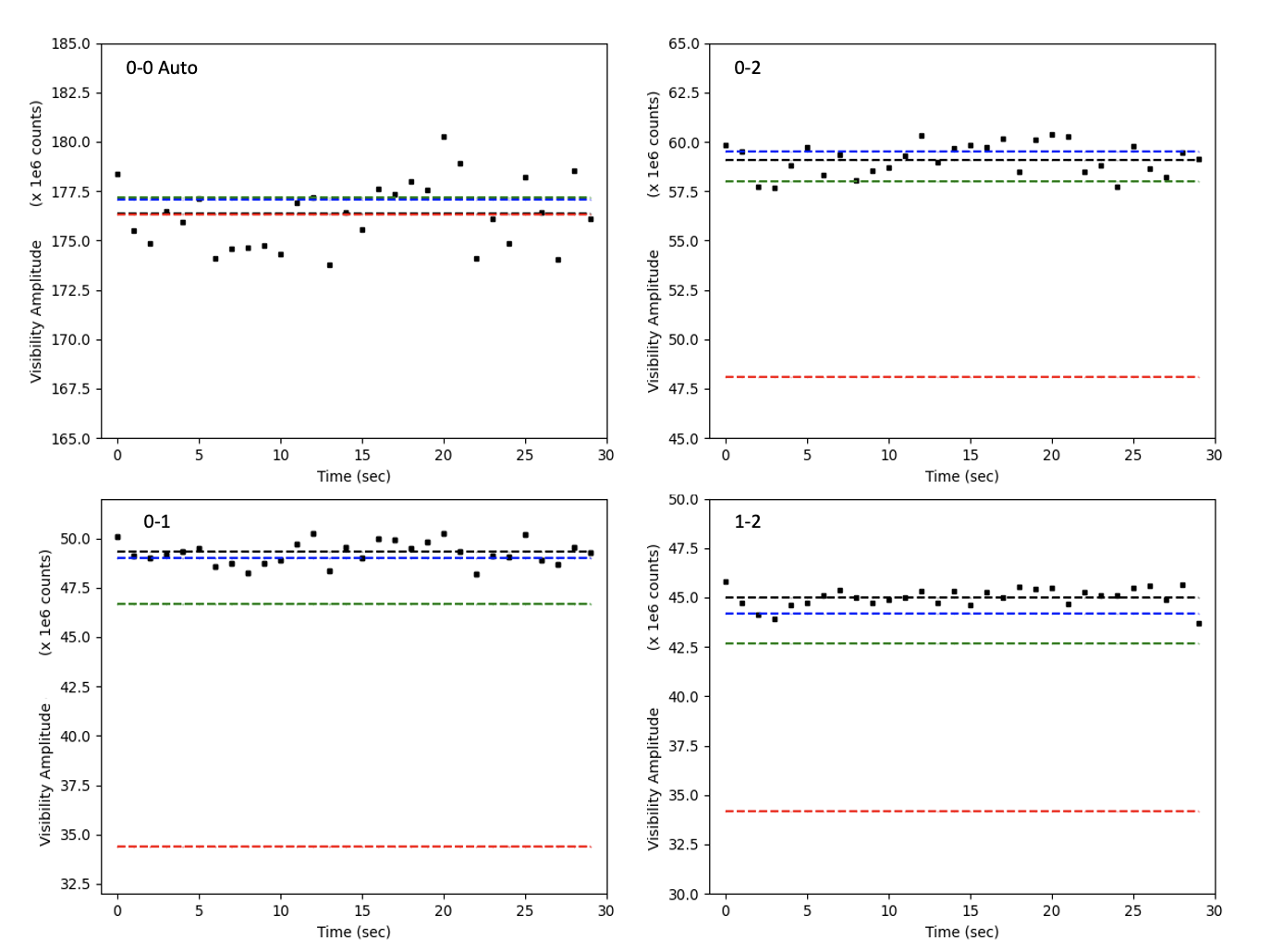}}
\caption{Visibility amplitudes for the four U,V data points of the 3-hole interferograms: (0,0) = autocorrelation; the other baselines for the uv-samples can be seen in Figure~\ref{fig:mask6H}. Black dots = results from the 30 individual frames. The black line is the average of these individual measurements, representing the target values for the IPSC process. The red line is after IPSC iteration 1 = summing images with no centering. The green line is after iteration 2 = summing images after Airy disk centering. The blue line is after iteration 3 = summing images after centering through cross correlation with the output from iteration 2. 
}
\label{fig:3Hamps}
\end{figure}

First, we adopt the 'target' value for the visibility amplitudes as those derived by taking the mean value of the results for the 30 individual 1\,ms frames (black points and line in Figure~\ref{fig:3Hamps}). Again, each frame has high S/N (many photons), and SOS conservation implies that the only change between frames is an overall-shift of the interference pattern, which will not alter the visibility amplitudes. 

Second, we derive visibility amplitudes after summing all the images with different alignments per frame, or centering, steps. For the 'worst case' (iteration 1), we simply sum all the images without centering, and derive visibility amplitudes from the Fourier transform of the summed image. Such a summed image will include decoherence due to the image wander caused by phase fluctuations. 

The second iteration of IPSC uses an Airy-disk centering process, in which each interferogram is smoothed to a low resolution corresponding roughly to the size of the Airy disk power pattern \cite{Nikolic2024} (see Figure~\ref{fig:movie}). The image shifts are derived by comparing the peaks of the smoothed images. This centering essentially removes the 'tip-tilt' term of the phase screen, ie. the mean phase slope across the full mask corresponding to the lowest order term in the phase screen. 

The third iteration of IPSC employs the output of iteration 2, meaning a model image corresponding to the sum of all the 30 Airy-disk centered images. Each 1\,ms frame is then spatially cross correlated with the model image to derive offsets, which are then removed and a new summed image is formed and analyzed. This third iteration would, assuming SOS conservation, remove higher order phase terms for 3-hole interferograms, beyond a uniform gradient, and thereby increase the coherence. 

Figure~\ref{fig:3Hamps} shows the time series of the visibility amplitudes from each frame, plus the mean value of this time series. Again, this mean represents the 'target' amplitudes for the ISPC test. For reference, the auto-correlation point is also included, corresponding to the total power in the field (0-0 visibility point). This autocorrelation barely changes with IPSC iteration, remaining unchanged to within 0.5\%, as expected since the total number of photons in the summed image in each iteration doesn't change with centering. 

For the interferometric visibilities (cross correlations), the 'no centering' summation shows the lowest amplitudes, lower by up to 30\% relative to the target values. The Airy disk centering increases the visibility amplitudes to within 5\% of target. The final IPSC iteration increases the amplitudes of the visibilities from summed images to within $\sim 1\%$ of the target values, well within the scatter of the points from the individual frames. 

We note that, while the source is only marginally resolved by the baselines of the uv-samples in the interferometric mask, the signal-to-noise of the measurements are very high, and the resulting visibility coherences provide reliable information on the emitting source size \cite{Torino2016, Nikolic2024}. The visibility amplitudes for the three uv-data samples employed herein are within $\sim 1\%$ of the same uv-data samples found in the 5-hole mask analysis by \cite{Nikolic2024}. \cite{Nikolic2024} show that, even though the implied coherences are high ($\ge 65\%$), the errors are small ($\le 1\%$), and a source size can be derived from Gaussian fitting to the visibilities. The fact that visibility amplitudes in the IPSC process for the 3-hole data converge to the target values, without resort to altering the phases of the uv-data itself, i.e. making corrections only in the image-plane, is then a direct confirmation of the IPSC process for interferometric imaging. 

\section{Summary}

We demonstrate, visually and quantitatively, Shape-Orientation-Size conservation using a 3-hole aperture mask interferometer at visible light wavelengths at the ALBA synchrotron light source. We also demonstrate the lack of SOS conservation for a 5-hole mask. 

We use the 3-hole mask interferograms to demonstrate the image plane self-calibration process. The demonstration entails monitoring the improvement in visibility amplitude with different image alignment iterations using improved source models. We find the process converges quickly, with Airy disk centering (tip-tilt) constituting the dominant, but not complete, phase correction. The final visibility amplitudes are very close to the target values seen in both the 3-hole and 5-hole masks, and hence provide direct information on the source size in the context of Gaussian fitting, as is done in \cite{Nikolic2024}.

A full demonstration of IPSC, including imaging of a complex source, requires a specialized interferometric setup, in which multiple triad interferograms are made in parallel, but separately, using beam splitters and waveguide optics, as described in \cite{CNT2023}.

{\bf Acknowledgments.} The National Radio Astronomy Observatory is a facility of the National Science Foundation operated under cooperative agreement by Associated Universities, Inc.. Image processing was performed using the Software: Astronomical Image Processing System (AIPS) \cite{Greisen2003} and Common Astronomical Software Applications (CASA) \cite{casa:2017}. We thank B. Kent for help with Figure 3. 

{\bf Disclosure Statement:} The authors declare no conflicts of interest.

{\bf Data availability:} The ALBA interferograms  underlying the results presented in this paper are not publicly available at this time but may be obtained from the authors upon reasonable request.

\clearpage
\newpage




\end{document}